\documentclass[9pt,twocolumn,twoside]{osajnl}

\journal{ol} % Choose journal (ao, ol, josaa, josab)

\setboolean{shortarticle}{true} % true = letter, false = research article

\title{Spectral symmetry of Fano resonances in a waveguide coupled to a microcavity}

\author[1*]{Andreas Dyhl Osterkryger}
\author{Jakob Rosenkrantz de Lasson}
\author[2]{Mikkel Heuck}
\author{Yi Yu}
\author{Jesper Mørk}
\author{Niels Gregersen}

\affil[1]{DTU Fotonik—Department of Photonics Engineering, Technical University of Denmark, Building 343, Kgs. Lyngby, Denmark}
\affil[2]{Department of Electrical Engineering and Computer Science, Massachusetts Institute of Technology, Cambridge, Massachusetts 02139, USA}

\affil[*]{Corresponding author: adyh@fotonik.dtu.dk}

\dates{Compiled \today}

\ociscodes{(050.2230) Fabry-Perot; (130.5296) Photonic crystal waveguides; (130.4815) Optical switching devices; (160.5298) Photonic crystals; (260.5740) Resonance}

\doi{\url{http://dx.doi.org/10.1364/ol.XX.XXXXXX}}

\begin{abstract}
We investigate the parity of transmission spectra in a photonic crystal (PhC) waveguide with a side-coupled cavity and a partially blocking element. We demonstrate, by means of numerical calculations, that by varying the cavity-block distance the spectra vary from being asymmetric with blue parity, to being symmetric (Lorentzian), to being asymmetric with red parity. For cavity-block distances larger than five PhC lattice constants, we show that the transmission spectrum is accurately described as the transmission spectrum of a Fabry-Perot etalon with a single propagating Bloch mode, and that the parity of the transmission spectrum correlates with the Fabry-Perot roundtrip phase.
\end{abstract}

\setboolean{displaycopyright}{true}

\begin{document}

\maketitle
\thispagestyle{fancy}
\ifthenelse{\boolean{shortarticle}}{\abscontent}{}

\begin{figure}[htbp]
\centering \def \svgwidth{\linewidth}
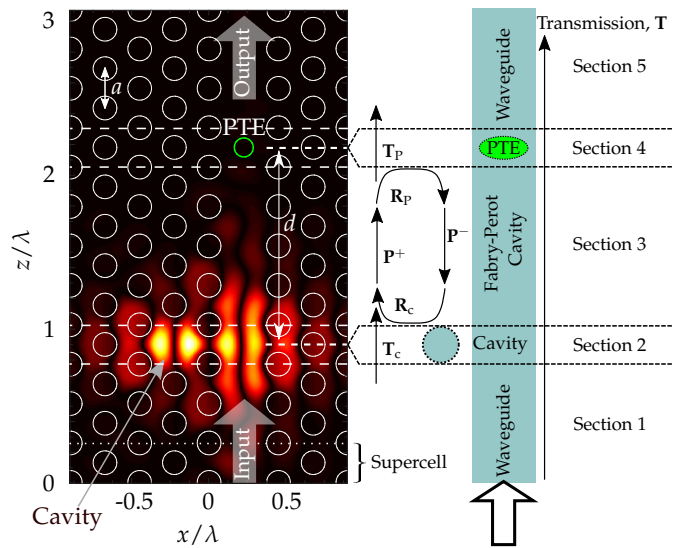
\caption{Left: PhC structure and field plot ($|H_y|$) at the minimum transmission frequency for the PhC Fano structure with hole radius $r = 0.30 a$, PTE radius $r_{\mathrm{PTE}} = 0.80 r$, Fabry-Perot length $d = 5 a$, refractive index of background material $n_\mathrm{b} = 3.1$ and refractive index of air holes $n_\mathrm{h} = 1$. The supercell for the first section is illustrated by the dotted white line, and the section interfaces are indicated with the dashed white lines. Right: Schematic of the structure with transmission, reflection and propagation matrices indicated, where the full PhC structure is divided into five sections.}
\label{fig:PhC_field_sketch}
\end{figure}

\begin{figure}[htbp]
\centering \def \svgwidth{\linewidth}
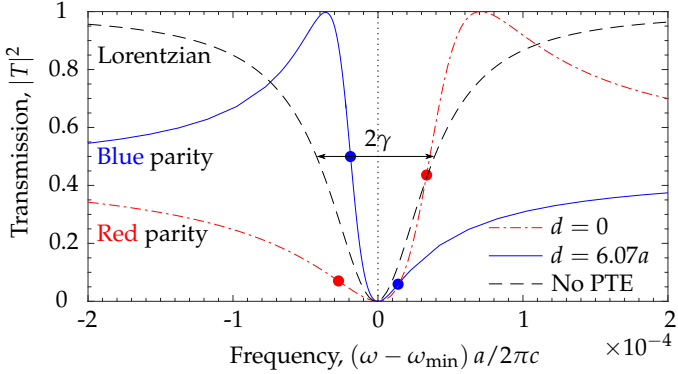
\caption{Transmission spectra for $d = 0a$ (\text{\color{red} red} parity), $d = 6.07a$ (\text{\color{blue} blue} parity) and for the microcavity only, where the geometry of the PhC structure is seen in Fig. \ref{fig:PhC_field_sketch}. The point with maximum slope on each side of the transmission minimum are indicated with solid markers for the two Fano spectra. The linewidth of the Lorentzian spectrum is $2\gamma$.}
\label{fig:Transmission_Parity}
\end{figure}

Photonic crystal (PhC) membrane structures consisting of waveguide-coupled microcavities represent an attractive platform for applications that can exploit the strong sensitivity of the transmission on the resonance frequency of the cavity. Due to the large ratio of quality factor to mode volume of PhC cavities \cite{Ryu2003}, even small refractive index perturbations within the volume occupied by the cavity mode leads to significant transmission changes. This fact has been used to demonstrate ultra-low energy all-optical signal processing~\cite{Nozaki2010} as well as chemical- and biological sensing \cite{Fenzl2014}. It was shown in 2002~\cite{Fan2002} how a Fano resonance~\cite{Fano1961} can be achieved in PhC structures, which further improves the wavelength sensitivity. The interference between a narrow- and broad-band state, which leads to Fano resonances, was implemented with a low- and high-$Q$ cavity structure for switching purposes~\cite{Nozaki2013}. We recently proposed a simpler geometry~\cite{Heuck2013ImprovedSwitching} and demonstrated that the shape of the transmission can be controlled~\cite{Yu2014FanoSwitching}. In this letter, we expand on these results by showing how both the parity and shape may be controlled in a way that is easily controlled experimentally. The geometry investigated in this letter is shown in Fig. \ref{fig:PhC_field_sketch}. We define the parity to denote whether the minimum of the transmission from the input to the output waveguide is red or blue shifted relative to the maximum, see Fig. \ref{fig:Transmission_Parity}. Different physical mechanisms cause the cavity resonance shift to be either positive or negative. In optical signal processing, depending on the preferred modulation format, it is essential whether the resonance shift causes an increase or decrease in transmission. Since this is determined by the parity of the resonance, our investigated structure is easily transferred between applications, where different signs of the resonance shift are demanded.

Figure \ref{fig:PhC_field_sketch} shows the investigated structure consisting of a microcavity adjacent to a waveguide containing a partially transmitting element (PTE), which was also a key element in previous proposals~\cite{Fan2002, Yu2014FanoSwitching, Heuck2013ImprovedSwitching}. By shifting the position of the PTE, both the parity and shape of the transmission spectrum can be controlled. The PTE is realized by a hole placed in the center of the waveguide and the microcavity is simply a point defect, i.e., a missing hole. The distance between the microcavity and the PTE is $d$, and $a$ is the PhC lattice constant. 

The structure investigated here is two-dimensional (invariant along $y$), and we use a Fourier-based Bloch mode expansion technique for simulating the transmission \cite{Lavrinenko2014nummethods, deLasson2014Blochmodal, deLasson2014RoundMat}. The structure is partitioned into periodic sections as shown in the right part of Fig. \ref{fig:PhC_field_sketch}, each with a distinct supercell and set of Bloch modes, and the expansions are coupled together with a Bloch mode S-matrix algorithm \cite{Lavrinenko2014nummethods}. Thus, we have direct access to the individual Bloch modes and their reflection and transmission coefficients, which plays a key role in the analysis to be presented here. The Bloch modes are determined in each section as in \cite{deLasson2014RoundMat, Lavrinenko2014nummethods} and the electromagnetic field is expanded on these Bloch modes:
\begin{equation}
\mathbf{H}_w(r) = \sum_{m} \left( a_{wm}\mathbf{\Psi}_{wm}^{H+}(\mathbf{r}_{\perp},z) + b_{wm}\mathbf{\Psi}_{wm}^{H-}(\mathbf{r}_{\perp},z) \right),
\end{equation}
\noindent
where $\mathbf{H}_w(r)$ is the magnetic field in the $w$th section and $a_{wm}$ [$b_{wm}$] is the amplitude of the $m$th forward ($+z$) [backward ($-z$)] propagating Bloch mode, $\mathbf{\Psi}_{wm}^{H+[-]}(\mathbf{r}_{\perp},z)$.

The transmission and reflection of the microcavity (PTE) section are computed by considering sections 1-3 (3-5), and using the scattering matrix formalism on this reduced geometry. This effectively reduces the full five-section geometry to a three-section geometry consisting of three waveguide sections (1, 3 and 5) coupled through the transmission and reflection matrices of the microcavity ($\mathbf{T}_\mathrm{c}$, $\mathbf{R}_\mathrm{c}$) and PTE ($\mathbf{T}_\mathrm{P}$, $\mathbf{R}_\mathrm{P}$) sections. Thereby, the total transmission from input to output in Fig. \ref{fig:PhC_field_sketch} is given as \cite{Lavrinenko2014nummethods}:
\begin{align}
\label{eqn:Trans_mat_eqn}
\mathbf{T} &= \mathbf{T}_{\mathrm{P}} \mathbf{P}^+\left(\mathbf{I} - \mathbf{RT} \right)^{-1} \mathbf{T}_{\mathrm{c}}, \\
&\mathbf{RT} \equiv \mathbf{R}_{\mathrm{c}} \mathbf{P}^{-} \mathbf{R}_{\mathrm{P}} \mathbf{P}^{+}, \label{eqn:RT_mat_eqn}
\end{align}
\noindent
where the matrices $\mathbf{P}^{+}$ and $\mathbf{P}^{-}$ represent propagation in section 3 by the length of an integer number of supercells in the forward and backward direction, respectively.  From Eq. (\ref{eqn:Trans_mat_eqn}) it is clear that the structure in Fig. \ref{fig:PhC_field_sketch} can be thought of as a Fabry-Perot cavity, where the microcavity constitutes a highly dispersive mirror, and this interpretation was previously used to propose an ultra-high speed laser structure~\cite{Mork2014Fanolaser}. When the mirror distance, $d$, is small enough for the PTE to lie inside the neighbouring supercell of the microcavity, the Fabry-Perot interpretation no longer makes sense, since this interpretation requires a waveguide supercell to be in-between the supercells of the PTE and the microcavity. In this case the structure will instead be divided into 3 or 4 sections (see Fig. \ref{fig:PhC_field_sketch}) and the total transmission matrix takes a different form.

\begin{figure}[htbp]
\centering \def \svgwidth{\linewidth}
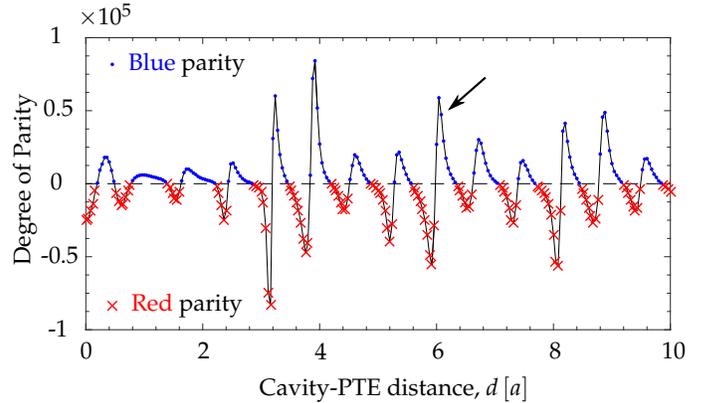
\caption{The degree of parity, $\mathrm{DoP}$, (defined in Eq. (\ref{eqn:DoP})) as function of the cavity-PTE distance, $d$. The data points are color-coded according to the parity and the black curve is a guide to the eye. The arrow indicates the DoP for the transmission spectrum with $d = 6.07a$ in Fig. \ref{fig:Transmission_Parity}.}
\label{fig:parity_vs_d}
\end{figure}

The transmission spectra for different cavity-PTE distances, $d$, are computed using Eq. (\ref{eqn:Trans_mat_eqn}) and a measure of the degree of parity, $\mathrm{DoP}$, is defined as the difference between the numerical maximum slope of the transmission spectrum before and after the transmission minimum (see the solid markers on the spectra in Fig. \ref{fig:Transmission_Parity}):
\begin{equation}
\label{eqn:DoP}
\mathrm{DoP} = \frac{2\pi c}{a} \left[ \max \left( \left|\frac{\partial T}{\partial \omega} \right|_{\omega < \omega_\mathrm{min}} \right) - \max \left( \left|\frac{\partial T}{\partial \omega}\right|_{\omega > \omega_\mathrm{min}} \right) \right]
\end{equation}
\noindent
With this definition, a positive (negative) DoP corresponds to blue (red) parity, and in Fig. \ref{fig:parity_vs_d}, the DoP is plotted for different cavity-PTE distances, where the points are color-coded according to the parity. It is apparent that the parity and shape of the transmission spectrum can be engineered by the position of the PTE relative to the microcavity, and very large slopes are achievable. An example of this is seen in Fig. \ref{fig:Transmission_Parity} with $d = 6.07a$, where the spectral distance between the maximum and minimum is seen to not be limited by the microcavity linewidth, $\gamma$, as is the case for our previously proposed structure with $d = 0$ \cite{Heuck2013ImprovedSwitching, Yu2014FanoSwitching}. A shorter spectral distance between the maximum and minimum can be obtained, but not while requiring $\max(|T|^2) = 1$ for our investigated structure.

The relative position of the transmission maximum and minimum results from the interference between many Bloch modes bouncing back and forth between the mirrors, as described by Eq. (\ref{eqn:Trans_mat_eqn}). Generally, it is not obvious how to determine the parity by direct inspection of this matrix equation. However, for sufficiently large $d$, the coefficients in $\mathbf{P}^{\pm}$ corresponding to evanescent modes are exponentially damped. For single-mode PhC waveguides, that we restrict the following analysis to, this means that only one element from the propagation matrices has a significant contribution and thereby all other elements can be neglected. This reduces the transmission Eqs. (\ref{eqn:Trans_mat_eqn}) and (\ref{eqn:RT_mat_eqn}) to scalar equations:
\begin{align}
\label{eqn:Trans_sca_eqn}
T =& T_{\mathrm{P}} P^+ \left(1 - RT \right)^{-1} T_{\mathrm{c}}, \\
RT &= R_{\mathrm{c}} P^- R_{\mathrm{P}} P^+, \label{eqn:roundtrip}
\end{align}
\noindent
where the (1,1) matrix elements are taken from the full matrices in Eqs. (2) and (3), since these couple and propagate the guided mode in the three waveguide sections (the same enumeration of the modes as in \cite{Lavrinenko2014nummethods} has been used).

\begin{figure}[htbp]
\centering \def \svgwidth{\linewidth}
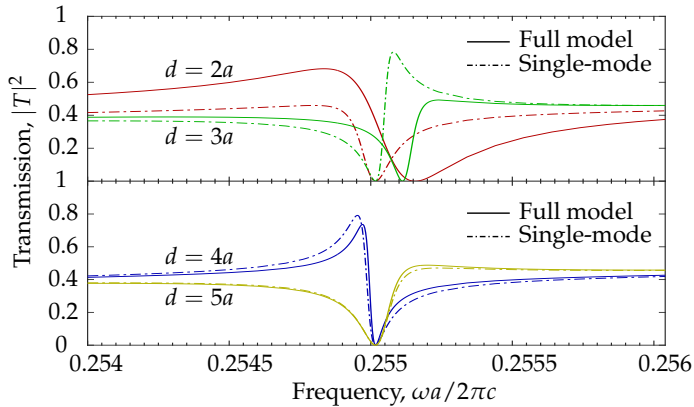
\caption{Transmission spectra for cavity-PTE distances $d = 2, 3, 4$ and $5a$ using Eq. (\ref{eqn:Trans_mat_eqn}) (full model, solid curves) and Eq. (\ref{eqn:Trans_sca_eqn}) (single-mode, dash-dotted curves).}
\label{fig:Transmission_All_vs_prop}
\end{figure}

In Fig. \ref{fig:Transmission_All_vs_prop}, the transmission spectra found from Eqs. (\ref{eqn:Trans_mat_eqn}) - (\ref{eqn:RT_mat_eqn}) (full model) and from Eqs. (\ref{eqn:Trans_sca_eqn}) - (\ref{eqn:roundtrip}) (single-mode) are compared for four different cavity-PTE distances. At the smallest distances (top panel), the single-mode model predicts the correct parity, but otherwise deviates visibly from the numerically exact spectra, e.g. with a clear offset on the spectral position of the transmission minimum. As the distance is increased to $d = 4a$ (bottom panel, blue curves), the agreement between the numerically exact and the single-mode model becomes substantially better, and at the largest distance considered here, $d = 5a$, (bottom panel, yellow curves) the agreement is almost perfect. The mismatch between the full and the single-mode model is due to the influence of evanescent Bloch modes in the Fabry-Perot region. A similar behavior was observed in \cite{Hugonin2007} in describing transmission  between a ridge waveguide and a slow light PhC waveguide, and in \cite{Sauvan2005} in analyzing PhC L$n$ cavities.

The minimum transmission frequency is shifted for $d = 2a$ and $d = 3a$ compared to $d = 4a$ and $d = 5a$ in Fig. \ref{fig:Transmission_All_vs_prop}, which does not seem intuitive, since the transmission of the guided Bloch mode through the microcavity section is zero at the resonance frequency of the microcavity for all $d \geq 2a$. However, the scattering of the guided Bloch mode at the microcavity section will populate evanescent Bloch modes in the Fabry-Perot section. For large Fabry-Perot lengths the population of the evanescent Bloch modes will vanish before reaching the PTE and no scattering will occur. But for small distances there will be a finite population of the evanescent Bloch modes at the PTE, where they will scatter and populate the guided Bloch mode in section 5, resulting in a finite overall transmission of the guided Bloch mode from section 1 to 5 at the resonance frequency of the microcavity. This effect causes the shift of the transmission minimum for structures with small cavity-PTE distances.

To render Eqs. (\ref{eqn:Trans_sca_eqn}) - (\ref{eqn:roundtrip}) more easily interpretable, we write the propagation constants and $T$- and $R$-coefficients as follows:
\begin{align}
P^+(\omega) &= P^-(\omega) = \exp \left(\mathrm{i} k(\omega) L \right), \\
T_{\mathrm{P}}(\omega) &= t_{\mathrm{P}}(\omega) \exp \left(\mathrm{i} \phi_{t,\mathrm{P}}(\omega) \right), \\
R_{\mathrm{P}}(\omega) &= r_{\mathrm{P}}(\omega) \exp \left(\mathrm{i} \phi_{r,\mathrm{P}}(\omega) \right), \\
R_{\mathrm{c}}(\delta) &= \frac{\gamma}{-\mathrm{i}\delta + \gamma} = \frac{\gamma}{\sqrt{\delta^2 + \gamma^2}} \exp \left(\mathrm{i}\phi_{r,\mathrm{c}}(\delta)\right), \label{eqn:Refl_Cavity} \\
T_{\mathrm{c}}(\delta) &= \frac{-\delta}{-\mathrm{i}\delta + \gamma} = \frac{-\delta}{\sqrt{\delta^2 + \gamma^2}} \exp \left(\mathrm{i}\phi_{t,\mathrm{c}}(\delta)\right), \label{eqn:Trans_microcavity}
\end{align}
\noindent
where $L$ is the distance between the microcavity and PTE sections, $k(\omega)$ is the dispersion of the guided Bloch mode in the PhC waveguide, $\phi_{t(r),\mathrm{P}}$ are the phases related to transmission and reflection at the PTE, $t_{\mathrm{P}}= |T_{\mathrm{P}}|$ and $r_{\mathrm{P}}= |R_{\mathrm{P}}|$ are the transmission and reflection amplitudes for the PTE, and $\delta = \omega - \omega_\mathrm{min}$ is the detuning. Finally $\gamma$ is half the linewidth of the transmission spectrum of the microcavity (see Fig. \ref{fig:Transmission_Parity}), which equals the coupling rate between the microcavity and the waveguide. The microcavity reflection phase is derived from Eq. (\ref{eqn:Refl_Cavity}) and the result is $\phi_{r,\mathrm{c}} = \arctan\left(\delta/\gamma\right)$. Using this and Eq. (\ref{eqn:Trans_sca_eqn}) we find:
\begin{align}
|T|^2 = & \frac{|T_{\mathrm{P}}|^2 |T_{\mathrm{c}}|^2}{1 + |R_\mathrm{P}|^2 |R_\mathrm{c}|^2 - 2 |R_\mathrm{P}| |R_\mathrm{c}| \cos \left( 2 k L + \phi_{r,\mathrm{P}} + \phi_{r,\mathrm{c}} \right)} \nonumber \\
=& \frac{t_P^2 \delta^2}{\delta^2 + (1 + r_P^2) \gamma^2 - 2 r_P \gamma \sqrt{\gamma^2 + \delta^2} \cos \left( \Phi_{RT} \right)}, \label{eqn:Trans_abs}
\end{align}
%
%=& \frac{t_P^2 \delta^2}{\left( \gamma^2 + \delta^2 \right)\left(1 + \frac{r_P^2 \gamma^2}{\gamma^2 + \delta^2} - \frac{2 r_P \gamma \cos \left( \Phi_{RT} \right)}{\sqrt{\gamma^2 + \delta^2}} \right)}
%
\noindent
where the frequency dependence of all parameters has been suppressed, and $\Phi_{RT} = 2 K \delta L + 2 k(\omega_\mathrm{min})L + \phi_{r,\mathrm{P}} + \arctan\left(\delta/\gamma\right)$ is the phase of the roundtrip as a function of detuning for a waveguide with linear dispersion, where $1/K$ is the group velocity. In the single-mode limit, the transmission vanishes exactly at the resonance frequency of the microcavity, i.e. at zero detuning $\delta = 0$, which is evident from Eqs. (\ref{eqn:Trans_microcavity}) and (\ref{eqn:Trans_abs}).

\begin{figure}[htbp]
\centering \def \svgwidth{\linewidth}
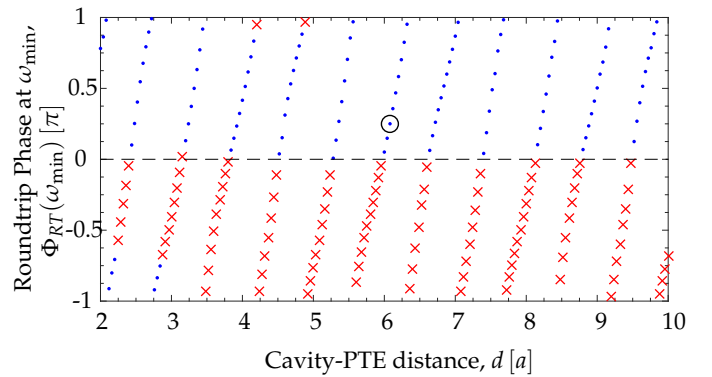
\caption{Phase of the roundtrip matrix element given in Eq. (\ref{eqn:roundtrip}) for different cavity-PTE distances. Each point has been color coded according to the parity of the numerically exact transmission spectrum (obtained from Eqs. (\ref{eqn:Trans_mat_eqn}-\ref{eqn:RT_mat_eqn})), such that {\color{red}red} parity structures are marked by {\color{red}red} crosses and {\color{blue}blue} parity structures are marked with {\color{blue}blue} dots. The black circle indicates the chosen cavity-PTE distance used for Fig. \ref{fig:Trans_different_RTMin}.}
\label{fig:PhiOneRoundTrip}
\end{figure}

Figure \ref{fig:PhiOneRoundTrip} shows the phase of the roundtrip element $RT$ in Eq. (\ref{eqn:roundtrip}) at the frequency of minimum transmission, $\omega_{\mathrm{min}}$, as a function of $d$. The blue (red) dots (crosses) correspond to the structure having blue (red) parity, where the parity is found from the full computation using Eq. (\ref{eqn:Trans_mat_eqn}). From our definition of parity in Eq. (\ref{eqn:DoP}) it follows that the transition between blue and red parity occurs when the transmission spectrum is an even function of the detuning, $\delta$. Eq. (\ref{eqn:Trans_abs}) shows that this can only be achieved, if $\cos(\Phi_{RT})$ is also even, which occurs when $\Phi_{RT}$ is odd, corresponding to $\Phi_{RT}(\omega_\mathrm{min})=0+p\pi, p\in \mathbb{Z}$. Since the transition only happens at these values, the parity must have the same sign in the intervals $\Phi_{RT} \in ]0; \pi[$ and $]{-\pi}; 0[$, which Fig. \ref{fig:PhiOneRoundTrip} confirms. The parity of the transmission spectrum is therefore completely determined by the roundtrip phase at the transmission minimum. 

However, the above explanation assumes that the transmission and reflection coefficients for the PTE, $t_\mathrm{p}(\omega)$ and $r_\mathrm{p}(\omega)$, are independent of frequency, which is generally not the case. This frequency dependence contributes to the asymmetry of the transmission spectra, but as seen in Fig. \ref{fig:PhiOneRoundTrip} the effect is very small, since the roundtrip phase at $\omega_\mathrm{min}$ predicts the right parity for all simulations with $d > 5a$. The above explanation assumes a linear dispersion, and thus a frequency independent group velocity. If, in turn, the structure is operated closer to the band edge of the waveguide, where the group velocity depends strongly on frequency, this would also affect the symmetry and could potentially be used as an additional knob to engineer the shape of the transmission spectrum.

\begin{figure}[htbp]
\centering \def \svgwidth{\linewidth}
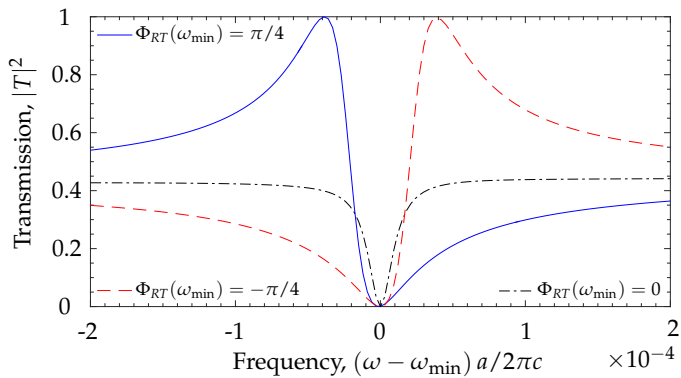
\caption{Transmission spectrum for $\Phi_{RT}(\omega_\mathrm{min}) = \pi/4$ ({\color{blue} blue} solid), $\Phi_{RT}(\omega_\mathrm{min}) = 0$ ({\color{black} black} dot-dashed) and $\Phi_{RT}(\omega_\mathrm{min}) = -\pi/4$ ({\color{red} red} dashed), where all other parameters are those for $d = 6.08a$.}
\label{fig:Trans_different_RTMin}
\end{figure}

Since the parity depends on the roundtrip phase, it is possible to flip the sign of the DoP by changing $\Phi_{RT}(\omega_\mathrm{min})$. This is shown in Fig. \ref{fig:Trans_different_RTMin}, where the transmission computed from Eq. (\ref{eqn:Trans_abs}) is plotted using the parameters for $d = 6.08a$ for $\Phi_{RT}(\omega_\mathrm{min}) = 0$ and $\pm\pi/4$. For this to be possible in an efficient way, the spectral distance between the maximum and minimum transmission points should be as small as possible. The investigated structure is not optimal, since it requires a total phase shift of $\sim \pi/2$ for switching the DoP and maintain $\max(|T|^2) = 1$. Reducing the linewidth, $2 \gamma$, and the PTE transmission, $t_\mathrm{PTE}$, would increase the slope and thus reduce the required phase shift for flipping the DoP, while maintaining $\max(|T|^2) = 1$.
\\

To conclude, we have analyzed the transmission spectrum of a photonic crystal microcavity coupled to a partially-blocked waveguide. It was shown that the structure displays Fano resonances and that the symmetry of these can be controlled by varying the distance between the microcavity and the partially transmitting element. For sufficiently large distances, a single-mode description is accurately describes the shape of the transmission spectrum, and in this limit it was shown that the phase of the roundtrip within the Fabry-Perot cavity determines the parity of the Fano resonance. This limit was identified to be at $d \simeq 5a$ for the investigated structure. The breakdown of the single-mode description for $d < 5a$ is due to the increasing influence of evanescent Bloch modes for smaller Fabry-Perot cavities. The possibility of fully tailoring the Fano resonance in photonic crystal microcavity-waveguide structures might find applications in, for example, optical signal processing and sensing.

\noindent
Our results suggest that the shape of the transmission can be made extremely sensitive to changes in the roundtrip phase. It is therefore interesting to investigate whether the structure is more susceptible to refractive index changes in the waveguide, rather than in the microcavity, which is conventionally used \cite{Nozaki2010, Fan2002, Nozaki2013, Heuck2013ImprovedSwitching, Yu2014FanoSwitching}.

%Especially that the slope of the transmission spectrum in principle can be infinite is interesting, since the parity of such structures can be changed by infinitesimal perturbations on the roundtrip phase. This could be realised by filling the PTE hole with a liquid or a gas with a refractive index that changes due to some foreign object (e.g. temperature, pressure) or by replacing the waveguide with an expandable polymer that reacts to chemicals and thereby changes the roundtrip phase.
\section*{Funding Information}
The project is funded by two Sapere Aude grants (ID: DFF – 4005-00370 and DFF: - 1323-00752) awarded by the Danish Research Council for Technology and Production as well as by Villum Fonden via the NATEC Center of Excellence.

%\section*{Acknowledgments}
%
%\textbf{Acknowledgment.}

% Bibliography

\pagebreak

\section*{Informational fifth page}
In this section, we provide full information on the references to assist reviewers and editors.

\end{document}